# Crossover of Magnetoresistance from Fourfold to Twofold Symmetry in $SmB_6$ Single Crystal, a topological Kondo insulator

Zengji Yue[1], Xiaolin Wang,[1*] Duanliang Wang,[2] Jiyang Wang,[2,3] Dimi Culcer,[4] and Shixue Dou[1]

[1]*Institute for Superconducting and Electronic Materials, University of Wollongong, North Wollongong, NSW 2500, Australia*

[2]*State Key Laboratory of Crystal Materials, Shandong University, Jinan, Shandong 250100, China*

[3]*School of Materials Science and Engineering, South China University of Technology, Guangzhou, 510641, China*

[4]*School of Physics, The University of New South Wales, Sydney 2052, Australia*

Topological Kondo insulators have been attracting great attention from the condensed-matter physics community due to their fascinating topological and strongly correlated properties. Here, we report angle-dependent *c*-axis magnetoresistance (MR) oscillations in a Kondo insulator, $SmB_6$ single crystal, in a magnetic field of up to 13 T rotated in the *ab*-plane. Four-fold symmetric MR oscillations are first observed above 8 K, which result from the four-fold ($C_4$) degeneracy of the bulk Fermi surface of $SmB_6$. With decreasing temperature down to 2.3 K, the $C_4$ symmetry of the MR oscillations gradually weakens and $C_2$ symmetry appears. This demonstrates a crossover from three-dimensional bulk states to two-dimensional surface states and implies the possible emergence of topological nematic states. Our experimental observations shed new light on the metallic surface states and nematic states in the Kondo insulator $SmB_6$.

KEYWORDS: magnetoresistance, fourfold symmetry, topological Kondo insulators, surface states

## 1. Introduction

Strongly correlated electron systems have been attracting great attention from the condensed-matter physics community due to their many exotic electronic and magnetic properties, such as high critical temperature (high-$T_c$) superconductivity, colossal magnetoresistance (MR), and electronic nematic states [1-3]. Samarium hexaboride ($SmB_6$) is well known as a strongly correlated electron system, in which the 4*f*-electrons hybridize with the conduction 5*d*-electrons and form a narrow Kondo energy gap, $E_g \approx 15$ meV [4-7]. The magnetic, optical, and thermal transport properties of this narrow-gap semiconductor have been extensively investigated in the last half-century [8-12]. The most puzzling issue is the lack of divergence in the resistivity at low temperatures, which has always been attributed to in-gap states [13-16]. Recently, there have been theoretical and experimental demonstrations that such in-gap states might be topological surface states [17-24].

Angle-resolved photoemission spectroscopy (ARPES) has demonstrated four-fold symmetric electron pockets near the Fermi level ($E_F$) [25-29]. Possible topological surface states were observed at the $\overline{X}$ and $\overline{\Gamma}$ points of the surface Brillouin zone *k*-space.

Magnetotransport is a helpful way to understand and explore the electronic states in strongly correlated systems and topological insulators [30-32]. The measurement of quantum oscillations is a powerful tool and widely used to determine the Fermi surface (FS) geometry and the electron dispersion in various metals [33-35]. Angle-dependent MR oscillations can be used to parameterize the two-dimensional (2D) in-plane (*ab*-plane) FS of metals [36-38]. Organic metals and high-Tc superconductors have been successfully characterized through such experimental measurements [39-42]. On the other hand, in layered compounds, a magnetic field rotated in the *ab*-plane can be applied as a valley valve to tune the contribution of each valley to the total interlayer conductivity. Such field-induced polarization of valleys has been observed in Bi and $Sb_2Te_3$ single crystals [43,44].

Here, we report the dependence of *c*-axis MR oscillations in $SmB_6$ single crystals on the angle, $\varphi$, of a magnetic field rotated in the *ab*-plane with respect to the *a*-axis. As the field rotates, the contribution of each electron pocket to the total conductivity is tuned and the polarized conductivity of the four electron pockets is achieved. With decreasing temperature, the *c*-axis angle dependence of the MR oscillations exhibits a gradual loss of four-fold ($C_4$) symmetry. Such a transition demonstrates a crossover from a bulk metal to a bulk insulator with metallic surface states. These experimental observations shed new light on the anomalous low-temperature conductivity of the Kondo insulator $SmB_6$.

## 2. Experimental Measurements

Single crystals of $SmB_6$ were grown by the conventional aluminum (Al) flux method. The sample was etched with acid to remove the leftover Al on the surface. Figure 1(a) displays the CsCl-type crystal structure of $SmB_6$. It can be seen that the Sm ions are located at the corner sites of the unit cell, while the $B_6$ octahedron is located at the body center of the cubic lattice. Figure 1(b) shows the bulk and surface Brillouin zone for $SmB_6$. Figure 1(c) shows an optical microscope photograph of a piece of barlike single crystal with a rectangular shape. Figure 1(d) displays a schematic view of the experimental configuration and the crystal axes. The $SmB_6$ sample is 2 mm in length. Measurements of the *c*-axis magnetotransport between 2.3 and 300 K were conducted by the four-probe method using a Quantum Design 14 T Physical Properties Measurement System (PPMS). A sample rotator was used to perform the MR measurements at different field orientations with respect to the current direction [Figure 1(d)]. Here, we define MR = $[(\rho_B-\rho_0)/\rho_0] \times 100\%$, where $\rho_B$ and $\rho_0$ are the resistivity in a magnetic field $B$ and in zero field, respectively. The applied AC bias current was 0.1 mA, directed along the *c*-axis, with the magnetic field rotated in the *ab*-plane.

Prior to the anisotropic MR measurements, the temperature dependence of the resistivity was measured. As shown in Figure 2(a), the *c*-axis electrical transport properties without an external field are characterized by different regimes that are correlated to various energy scales. Above 40 K, the Kondo gap is closed, and the $SmB_6$ behaves as a metal. Between 40 and 3.5 K, the resistivity increases by several orders of magnitude due to the opening of the Kondo gap. Below 3.5 K, the resistivity approaches a constant value of approximately 5.3 Ω·cm. Figure 2(b) shows the magnetization as a function of temperature. The $SmB_6$ single crystal demonstrates paramagnetic behavior at high temperatures. The downturn in the magnetization at low temperatures can be attributed to antiferromagnetic correlations and instability due to the strongly correlated electron interaction at low temperatures [45,46]. Actually, there is competition among the topological, Kondo-coherent, and magnetic states in $SmB_6$, although the topological phase remains stable in the presence of the antiferromagnetic order for a range of values [47]. The surface states in $SmB_6$ also remain gapless unless the bulk gap closes. Figure 2(c) presents a series of measurements of the *c*-axis MR as a function of the field at different temperatures with $\varphi = 0^{\circ}$. The high-field negative MR in $SmB_6$ has been attributed to the field dependence of the gap through adjustment of the hybridization between the 4*f*-band and the 5*d*-band by the magnetic field [48-50]. Although this model could qualitatively explain the experimental observation, the detailed mechanism needs more theoretical investigation.

Figure 3 displays the evolution of angle-dependent MR oscillations at different temperatures at 13 T. On rotating the sample within the *ab*-plane, the angle-dependent *c*-axis MR oscillations have a near four-fold symmetry with peaks at approximately 45, 135, 225, and 315°. MR oscillations are detectable in a magnetic field of $B > 4$ T. As shown in Figure 3(a)-(c), fourfold *c*-axis MR oscillations can be clearly seen at 15, 10, and 8 K in a field of 13 T in the *ab*-plane. The four electron pockets are interchangeable and the four-fold symmetry of the $SmB_6$ lattice is maintained. The fourfold oscillations were expected because the bulk state of $SmB_6$ is a Kondo insulator and is four-fold degenerate in the ground state [25,26]. On decreasing the temperature to 5 K, nearly all the four-fold MR oscillations gradually disappear, as shown in Fig. 3(d). Simultaneously, another two unequal peaks at approximately 0° and 180° gradually appear and grow stronger as the temperature is reduced to 2.3 K. In addition, we tried the measurements of the angle-dependent MR in different magnetic fields, and found that the $C_4$ and $C_2$ symmetries are not evident below 4 T [see Figure 3(g)]. Also the high magnetic field 13 T was best for observing the transition from $C_4$ and $C_2$ symmetry.

## 3. Results and Discussion

Firstly, we attempt to use an empirical formula to fit our experimental data. In multiple-electron pocket systems, in the presence of an magnetic field in *ab*-plane, the *c*-axis conductivity ($\sigma$) is expected to be the sum of the contributions by individual electron pockets. In the case of the

Kondo insulator $SmB_6$, the conductivity can be written as

$\sigma = \sum_{i=1-4} \sigma^i$,

(1)

where the conductivities of the electron pockets are indexed by $\sigma^i$. Our experimental data can also be fitted by the empirical formula,[43,44]

$$\sigma = \sum_{i=1-4} \frac{\sigma^i{}_{Max}}{1+A\cos^2(\varphi+(i-1)\frac{\pi}{2})} . \quad (2)$$

In this formula, the contribution of each of the four electron pockets is described by the same angle-dependent function rotated by $\pi/2$. $A$ is a given parameter and represents the anisotropic oscillation of conductivity. Since one electron pocket is located in each of the four directions in *k*-space, $\sigma^i{}_{Max}$ is the maximum $\sigma^i$ when the direction of the magnetic field in the *ab*-plane becomes parallel to this axis. As the field rotates, the conductivity decreases and reaches a minimum (resistivity increases and reaches the maximum) when the magnetic field is oriented in a direction between the [100] axis and the [010 axis in *k*-space, where electron pockets are located on these axes. The angle-dependent MR is qualitatively consistent with Eq.(2), as shown in Figure 3 (a)-(c).

Secondly, we argue that the angle-dependent MR $C_4$ symmetry above 5 K can be attributed to the symmetry of the FS composed of the surface state. Figure 4 displays a schematic plot of the 2D FS of the Kondo insulator $SmB_6$ with five electron pockets. The 2D FS originated from the metallic surface states of the Kondo insulator $SmB_6$. On the basis of first-principles calculation of the band structure, they originate from the *d*-*f* hybridization with the 5*d*-derived states crossing the flat 4*f*-bands in $k_{x,y}$-space [25,26]. On the basis of ARPES experiments, the four electron pockets enclose the $\bar{X}$ point in the first Brillouin zone [25-28]. Only one isotropic electron pocket is located at the $\bar{\Gamma}$ point.

According to the results of qualitative analysis, the *c*-axis resistivity can be set using $\omega_c\tau$ [37,43,44]. Here $\omega_c\tau = \frac{eB\tau}{m_\perp}$, with $\omega_c, \tau, e,$ and $m_\perp$ corresponding to the cyclotron frequency, scattering time, charge, and the *c*-axis effective mass of the electron, respectively. A theoretical calculation also indicates that the dispersion crossing the Fermi level $E_F$ is considerably anisotropic around the $\bar{X}$-point [51]. The anisotropic dispersion suggests the anisotropy of the electron effective mass around $\bar{X}$-points. Quantum oscillation experiments and Landau spectrum also reveal the anisotropic electron mass in $SmB_6$ [52]. The calculated effective masses for electron pockets $\alpha$ and $\beta$ are 0.074 $m_e$ and 0.10 $m_e$, where the $m_e$ is the bare electron mass. These values play an important role in determining the transport properties of $SmB_6$.

Due to the *ab*-plane anisotropy, a polarized current can be expected in a well-oriented magnetic field. The mobility $\mu$ of *c*-axis electrons in the perpendicular planes determines their response to the magnetic field. These electrons with minimal cyclotron mass $m_\perp$ and maximum $\omega_c$ give rise to the largest MR [43,44]. Since different electron pockets respond differently to the orientation of the magnetic field, any two of the four electron pockets dominate the total conductivity, thereby creating the polarization of electron pockets. Hence, four-fold $\rho_c$ oscillations are expected in such a system with four electron pockets of light electrons.

The $C_2$ symmetry of the angle-dependent MR was observed in the temperature range where a resistance plateaus appears. In such a range, it is widely believed that the Kondo insulator $SmB_6$ has an insulating bulk with a Kondo gap and 2D metallic surface states [21-24]. Therefore, the transition of MR from $C_4$ to $C_2$ symmetry indicates a crossover from 3D metallic bulk states to 2D metallic surface states with an insulating bulk. Very recently, we noticed that a theoretical model has been established to explain our experimental observations [53]. This theoretical model demonstrates that the $C_2$ symmetry of the angle-dependent MR in $SmB_6$ may originate from the nematic surface states on the FS [53].

An electronic nematic state is a translationally invariant metallic phase with a spontaneously generated spatial anisotropy [3,54]. Such a state has been observed in strongly correlated systems including $URu_2Si_2$, $Sr_3Ru_2O_7$, and superconductors [3,54-57]. In the topological Kondo insulator $SmB_6$, strong interactions may drive this system into rotational-symmetry-breaking nematic states [53]. In addition, the anisotropy in the FS of $SmB_6$ may also promote the formation of nematic states at low temperatures. The nematic phase breaks the $C_4$ rotational symmetry and generates the $C_2$ symmetry in the angle-dependent MR. In addition, as shown in Figure 3(f), the angle-dependent MR below 3 K is not strictly symmetric. The differences in the MR values at 0 and 180° may result from the difference in the electron pockets that determine the surface conductivity.

## 4. Conclusion

In summary, we first experimentally observed angle-dependent *c*-axis MR oscillations in $SmB_6$ single crystal, which is a topological Kondo insulator, in a magnetic field rotated in the *ab*-plane. The fourfold symmetric MR oscillations are attributable to the $C_4$ degeneracy of the FS electron pockets. The low-temperature behavior suggests the broking of rotational symmetry and can be attributed to the appearance of nematic surface states in the Kondo insulator $SmB_6$. Our observations provide new experimental evidence to reveal the puzzle of low-temperature residual conductivity in the Kondo insulator $SmB_6$.

## Acknowledgements

Xiaolin Wang acknowledges support from the Australian Research Council under a Discovery project (ARC Discovery, DP130102956). This work was also partially supported by the Chinese National 973 program (2010CB833103).

## Figures

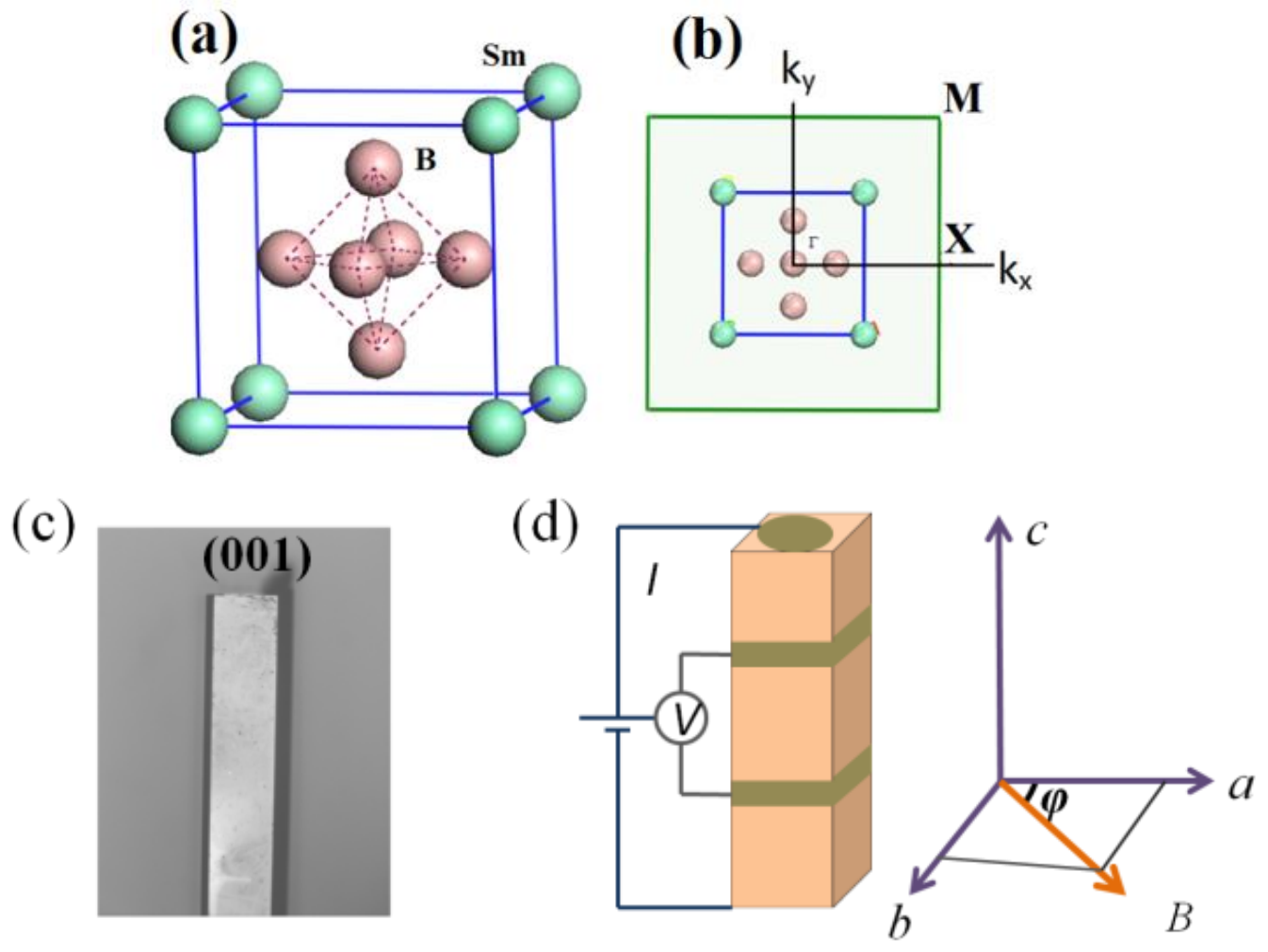

Figure 1 (a) CsCl-type crystal structure of $SmB_6$. (b) Bulk and surface Brillouin zone for $SmB_6$. (c) Photograph of $SmB_6$ single crystal. (d) Schematic view of experiment configuration. (Color online)

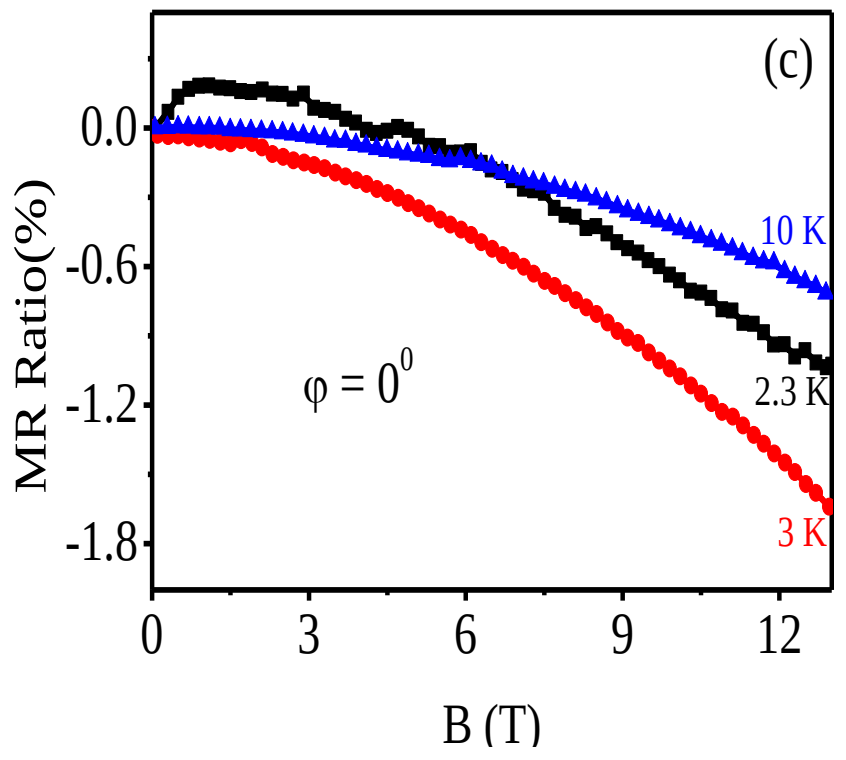

Figure 2 (a) Temperature dependence of the zero-field resistivity of the $SmB_6$ single crystals. (b) Temperature dependence of magnetization for $SmB_6$ single crystal. (c) Magnetic field dependence of the *c*-axis MR at different temperatures. (Color online)

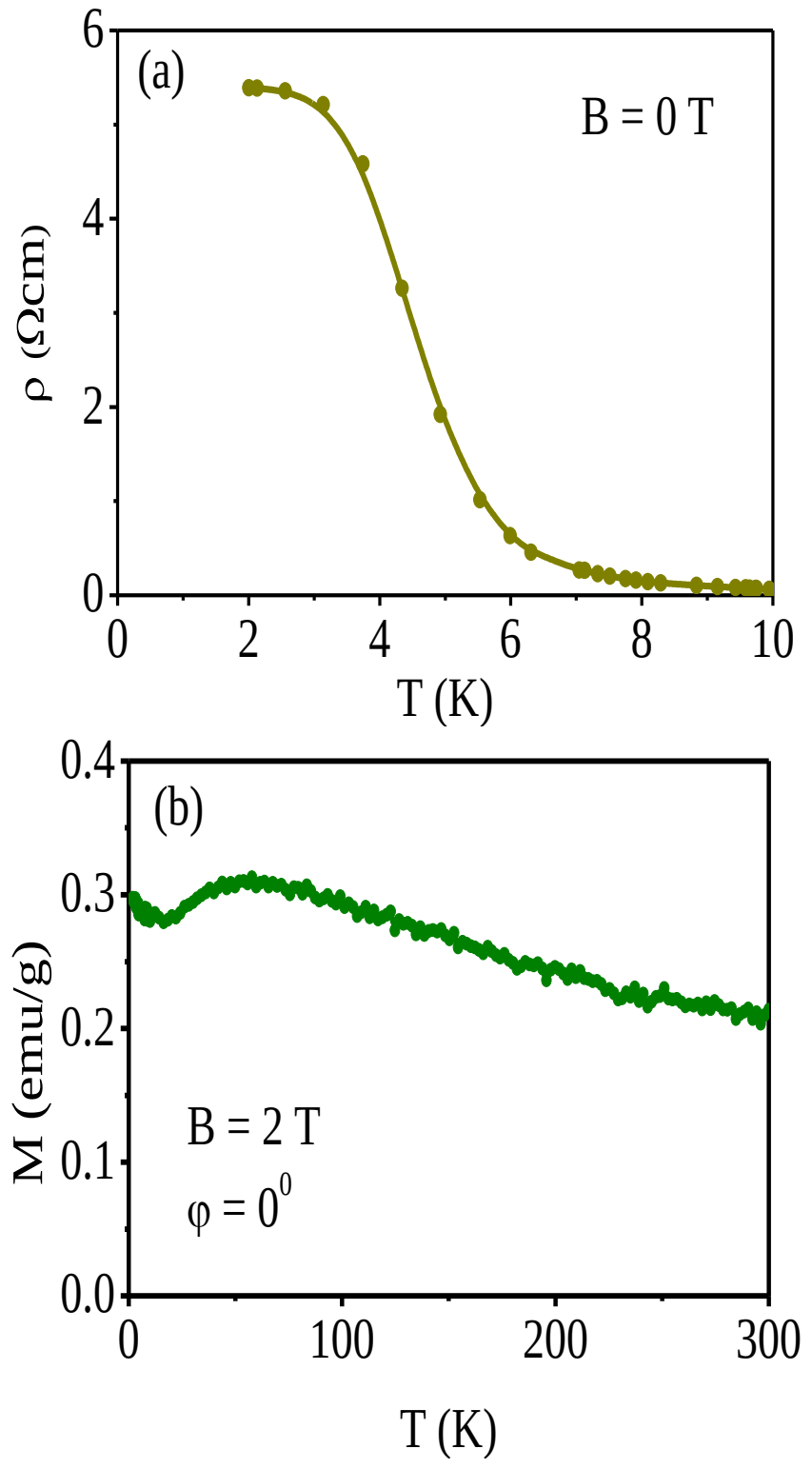

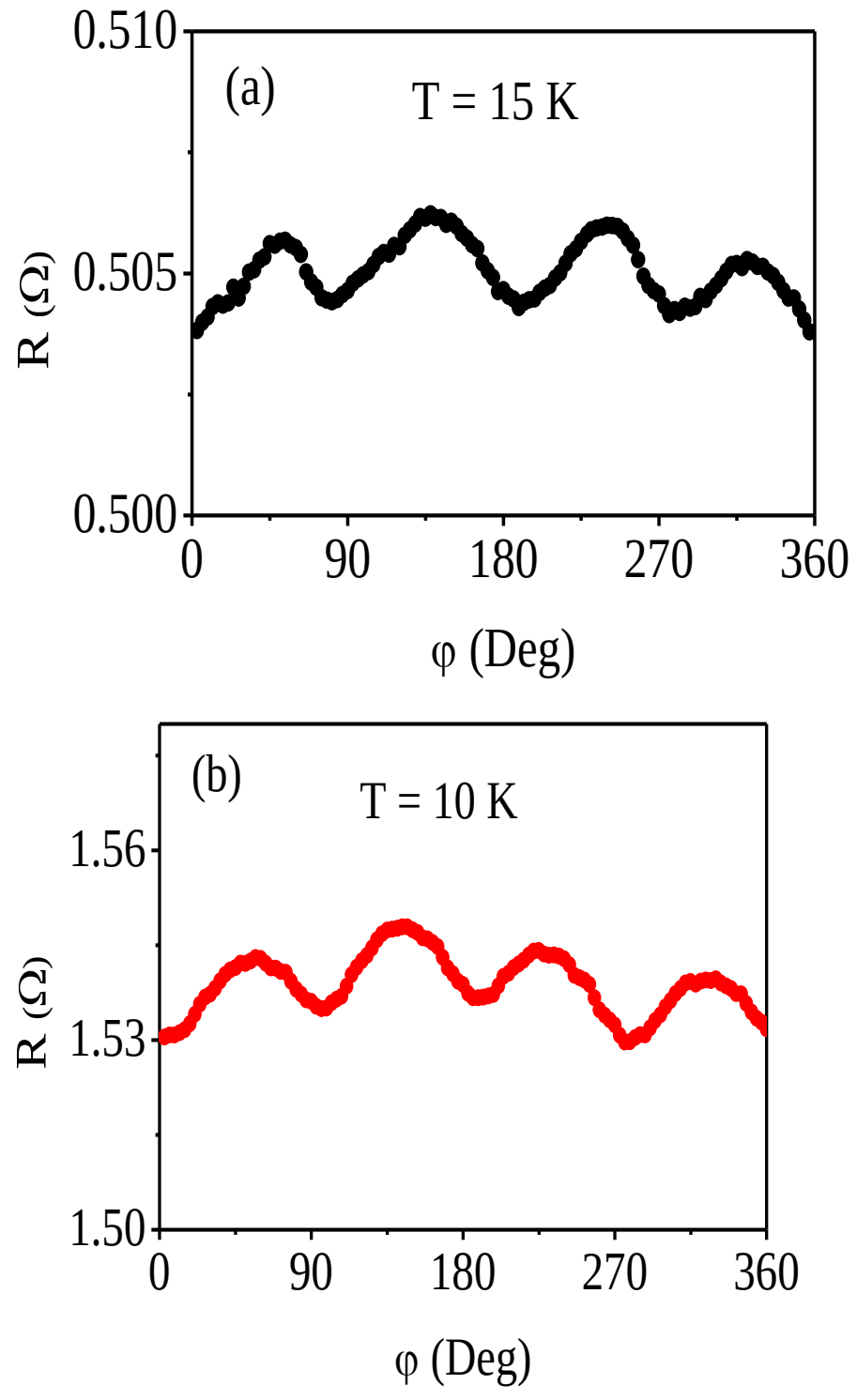

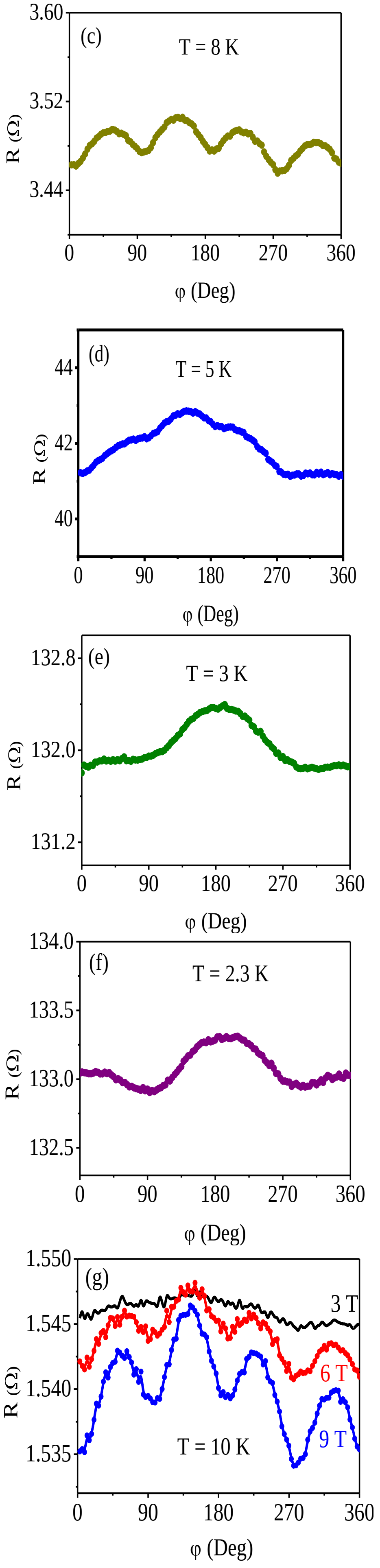

Figure 3 Angle-dependent *c*-axis MR oscillations in 13 T at different temperatures ranging from 15 K down to 2.3 K: (a) 15 K, (b) 10 K, (c) 8 K, (d) 5 K, (e) 3 K, (f) 2.3 K. (g) Angle-dependent *c*-axis MR oscillations at 10 K in different magnetic fields ranging from 3 T to 9 T. (Color online)

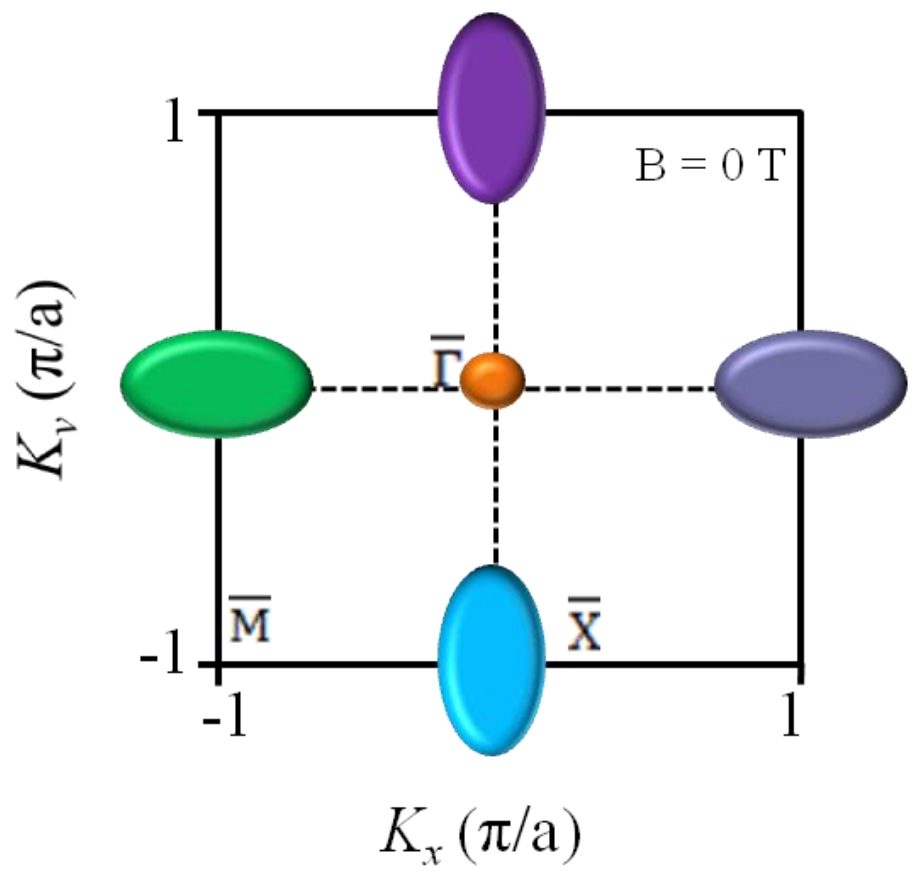

Figure 4 Schematic plot of the 2D FS of $SmB_6$ with four electron pockets in the first Brillouin zone. (Color online)